# Giant ultrafast optical enhancement of strain gradient in ferroelectric thin films

Yuelin Li[1*], Carolina Adamo[2], Pice Chen[3], Paul G. Evans[3],

Serge M. Nakhmanson[4], William Parker[5], Clare E. Rowland[6], Richard D. Schaller[7],

Darrell G. Schlom[8,9], Donald A. Walko[1], Haidan Wen[1], and Qingteng Zhang[3]

[1] *Advanced Photon Source, Argonne National Laboratory, Argonne, Illinois 60439, USA*

[2] *Department of Applied Physics, Stanford University, Stanford, CA 94305, USA*

[3] *Department of Materials Science and Engineering & Materials Science Program, University of Wisconsin–Madison, Madison, Wisconsin 53706, USA*

[4] *Department of Materials Science & Engineering, and Institute of Materials Science University of Connecticut, Storrs, CT 06269-3136, USA*

[5] *Argonne Leadership Computing Facility, Argonne National Laboratory, Argonne, Illinois 60439, USA*

[6] *Department of Chemistry, Northwestern University, Evanston, IL 60208, USA*

[7] *Center for Nanoscale Materials, Argonne National Laboratory, Argonne, Illinois 60439, USA*

[8] *Department of Materials Science and Engineering, Cornell University, Ithaca, New York 14853, USA*

[9] *Kavli Institute at Cornell for Nanoscale Science, Ithaca, New York 14853, USA*

**The coupling between strain gradients and polarization, known as flexoelectricity, offers a new mechanism to control the functionality of dielectric materials. However, for the effect to be practically attractive, dynamic control of the strain gradient with magnitudes far exceeding those achievable via mechanical deformation (~10 m$^{-1}$) is needed. Strain-engineered thin films exhibit extraordinary strain gradients of $10^5$-$10^6$ m$^{-1}$ arising**

**from structural relaxation within a short space range that greatly enhances the steady-state flexoelectric effect. Here we report a giant, optically initiated dynamic enhancement of the strain gradient, also on the order of $10^5$-$10^6$ m$^{-1}$, in ferroelectric BiFeO$_3$ epitaxial thin films via time-dependent coherence analysis of X-ray diffractions. The finding opens the door for dynamic coupling of the flexoelectric effect with light, making optical switching of polarization, and thus application such as direct optical writing of non-volatile ferroelectric memory, possible. A combination of time-resolved X-ray scattering and optical spectroscopy shows that the enhancement of the strain gradient is due to a piezoelectric effect driven by a transient screening electric field, opening the opportunity for new ways of studying flexoelectric effect in strain engineered ferroelectric thin films.**

Flexoelectricity can be used to control the direction and magnitude of the spontaneous ferroelectric polarization using the electric field resulting from a strain gradient field, termed the flexoelectric field[1]. However, flexoelectric control of the polarization has been limited to static conditions because the strain gradients generated by epitaxial relaxation[2–4] in epitaxial thin films[4–8] or by mechanical deformation[1,9] cannot readily be varied. The interaction of light with correlated materials has already generated rich phenomena important for new material functionalities and understanding the mechanisms governing these functionalities. Light excitation of ferroelectric complex oxides, in particular, has generated a plethora of intriguing and potentially useful yet largely unexplained physical phenomena including photostriction[10–12] and photovoltaic[13,14] effects. Among them, epitaxial multiferroic BiFeO$_3$ (BFO) [15] thin films have a strong structural and electronic response to excitation by photons with an energy larger than the direct band gap of 2.6-2.7 eV[16]. Absorption of these photons generates photo voltages larger than the bandgap[13,14] and produces large lattice distortions[12,17]. BFO epitaxial films also exhibit strain gradients larger than $10^5$ m$^{-1}$ due to structural relaxation[3] and a strong dependence of the polarization on flexoelectricity fields[7,8]. Here we demonstrate that the strain gradient in BFO thin films can be enhanced via ultrafast optical excitation by a magnitude comparable to the static strain gradient, i.e., $10^5$-$10^6$ m$^{-1}$, effectively doubling or tripling the flexoelectric field. This opens the possibility of direct coupling of flexoelectricity with optical stimuli and thus a new horizon in studying the mechanism of flexoelectricity and its applications such as direct optical writing of ferroelectric memories[18].

The varying strain profile due to strain gradients is readily measurable because the strain gradient imposes a spatially varying phase of the coherently scattered X-rays, thus changing the distribution of diffracted intensity in reciprocal space. The real-space variation of the strain can be reconstructed from an coherence analysis of the diffraction pattern[19] in combination with structural assumptions, for example, of the orientation of the strain gradient expected from strain relaxation. Time-resolved coherence diffraction methods, with a schematic shown in Fig. 1(a), enable the reconstruction of a spatiotemporal map of the strain with a time resolution limited only by the X-ray pulse duration.

Time dependent diffraction shows that the optical pump pulses produce significant changes in the strain and strain gradient in the BFO thin film. A comprehensive study on fluence and film thickness dependence revealed systematic variations in the magnitude and timescales of the structural transient. The samples in the experiment are phase-pure 4, 20 and 35 nm thick films grown on SrTiO$_3$ (STO). X-ray reciprocal space mapping[20] shows that the 4 nm film is tetragonal and the others are monoclinic. The distribution of X-ray intensity along a crystal truncation rod cut through the 002 BFO Bragg peak at different delays between a pump laser pulse and a probe X-ray pulse is shown in Fig. 1 (b) for the 35 nm film. More data are shown in the Supplementary Information (SI).

The fitted strain derived from a coherence analysis of the diffraction results is reported in Fig. 1(c) relative to the average out-of-plane pseudocubic lattice parameter of the 35 nm film, 4.074 Å. The coherence analysis, described in the SI, adjusts the strain profile, i.e., the relative phases of the diffraction unit cells, to fit the diffraction amplitude of the film to the experimentally measured data. Though highly nonlinear strain profiles are allowed, the fits for all time delays converge to ones that are dominated by a linear term (Figs. 1, SI Figs. S1, S2, and S3). The static strain profile without laser excitation exhibits a relative compressive strain near the free surface and an expansion in layers at the substrate-film interface, as shown in Fig. 1(c). This variation arises from relaxation of the epitaxial strain via polarization domain formation[21]. In line with other studies[4], the static strain gradients are $1.7 \times 10^5$/m and $3.9 \times 10^5$/m for the 35 and 20 nm films. For times after optical excitation, these fit strain profiles can be approximated by a linear relationship

$$\varepsilon(t, z) \approx \alpha(t)\varepsilon_0(z) + \beta(t) . \qquad (1)$$

From which, the strain gradient as a function of time is:

$$\frac{d\varepsilon(t,z)}{dz} \approx \alpha(t) \frac{d\varepsilon_0(z)}{dz}. \qquad (1a)$$

Here $\varepsilon_0(z)$ is the static strain profile before the arrival of the laser pulse while the strain after the excitation is tilted by the factor $\alpha(t)$ with a shift of $\beta(t)$. Direct comparison with the diffraction scans verifies this linear model, identifying the physical meaning of $\alpha$ and $\beta$ as follows: the tilting factor $\alpha(t)$ is proportional to the overall spread of the lattice distortion, which manifests itself in a broadening of the diffraction peak $\Delta w(t)$ (Fig. 1 (d)) because the broadening is due to the overall spread of the lattice distortion; the $\beta(t)$ term corresponds to the average transient strain $\Delta\varepsilon$ as derived from the angular shift of the Bragg peak (Fig. 1 (e)). Fitting at a different fluence and for the 20 nm film generates similar results (SI Figs. S2, S3). Remarkably, upon the optical excitation at around 3.3 mJ/cm$^2$, the strain gradient increases from $1.7\times10^5$/m and $3.9\times10^5$/m to $2.5\times10^5$/m and $1.0\times10^6$/m for the 35 and 20 nm films, respectively (Figs. 1(d), SI Fig. S2(b), and S3(b)).

To understand the mechanism driving the strain gradient change, the film thickness dependence was studied by varying the nominal absorbed laser fluence ranging from 0.6 to 2.5 mJ/cm$^2$ for all three samples. The broadening $\Delta w$ and the strain $\Delta\varepsilon$ of the (002) diffraction peak were measured as a function of the delay between the laser pump and the x-ray probe (Fig. 2). They represents the decoherence anf the average phase shift of the diffraction unit cells. We note that the recovery is smooth and monotonic, showing no local minima. The 1/$e$ recovery time for both $\Delta w$ and $\Delta\varepsilon$ is found to be independent of the fluence but becomes longer as the film becomes thicker, as shown Fig. 2 and summarized in Fig. 4. In addition, at the same fluence, the maximum $\Delta w$ and $\Delta\varepsilon$ are bigger for thinner films, e.g., at a fluence of about 2.5 mJ/cm$^2$, the maximum $\Delta w$ is 0.027, 0.017, and 0.011 degrees for the 4, 20, and 35 nm samples, with corresponding $\Delta\varepsilon$ of 0.7%, 0.6%, and 0.4%, respectively. For the same film thickness, $\Delta w$ and $\Delta\varepsilon$ exhibits a nearly linear dependence on the fluence.

Similarly, the relaxation of the photo-induced absorption in a transient absorption spectroscopy (TAS) experiment, measured as the photo-induced optical density (OD), also depends strongly on the thickness (Fig. 3). The 1/$e$ recovery time of the optical and structural relaxation are compared in Fig. 4. The purity of the sample phase excludes this thickness dependence from being the result of carrier trapping by dislocation or defects inside the films. In this case, as the decay of the OD is due to the removal of the carriers responsible for the absorption, the thickness dependence thus indicates that the carrier lifetime is determined by diffusion rate and surface annihilation[22]. The diffusion coefficient can be estimated[22] by fitting the time constant to a relation $\tau_{OD} = (Z/\pi)^2/D$, where $Z$ is the thickness of the films, giving an average diffusion coefficient $D=0.4 \pm 0.05$ nm$^2$/ps.

The optically induced strain has been identified as originating from photoexcited carriers with a slower thermal contribution from heating[12]. The effect had been interpreted as arising from the piezoelectric effect[11,12] due to the screening field formed by free carriers driven to the surface and interface by the internal polarization field. More recently, however, a model based on localized lattice distortion or dephasing by exciton formation[17], arising from inhomogeneous photo-deposition, has been proposed to explain the instantaneous onset of both the strain and broadening of the diffraction peak, an effect not expected from the above piezoelectric model due to the finite time it takes for the carriers to reach the surface and interface.

However, the localized carrier model is inconsistent with the thickness dependence data. The photo-deposition as a function of the depth $z$ into the film follows a simple exponential function, i.e., $\exp(-z/L)$, where $L = 32$ nm is the BFO absorption length for 400 nm light. As the film thickness becomes smaller, the photon deposition becomes more homogeneous, and thus the expected net broadening should reduce. This is contrary to our observation discussed above. Furthermore, as the deposition profile would induce a strain gradient with an opposite sign to the static strain gradient, the model also predicts a narrowing of the diffraction peak (of about 0.015 and 0.005 degrees maximum for the 20 and 35 nm films, estimated using the measured static strain) when the photo-effect roughly cancels the intrinsic strain gradient and, as such, a fluence-dependent recovery dynamics as well. This narrowing is absent from the delay and the fluence dependence measurement. Finally, the localization of the carrier also requires that the carriers recombine locally. Local carrier recombination implies a recovery dynamics that is independent of the film thickness, yet the opposite was observed. In addition, our independent DFT simulations also show no evidence of significant carrier-lattice correlation (SI Fig. S4).

To construct a physics model consistent with the experiment observation, we note that the distortion of the lattice in ferroelectric material arising from an applied electric field is connected by the piezoelectric coefficient $d_{33}$. The strain $\varepsilon$ after the application of the field is connected to that static strain $\varepsilon_0$ by

$$\varepsilon = \varepsilon_0 + d_{33}E. \tag{2}$$

Though generally regarded as a constant, $d_{33}$ macroscopically is dependent on the applied field[23] and epitaxial strain[24]. With Eq. (2) in mind, Eq. (1) can be rewritten as

$$\varepsilon(t,z) \approx \varepsilon_0(z) + \gamma\varepsilon_0(z)d_0E_{scr}(t) + \beta(t). \tag{3}$$

with the following relationships:

$$d_{33}(z) = [\gamma\varepsilon_0(z) + 1]d_0, \tag{4a}$$

$$\beta(t) = \varepsilon_h(t) + E_{scr}(t)d_0, \tag{4b}$$

$$\alpha(t) - 1 = \gamma d_0 E_{scr}(t). \tag{4c}$$

Here, $\gamma$ represents the linear dependence of the piezoelectric coefficient on the strain and $d_0$ is the average piezoelectric coefficient of the film, $E_{scr}$ is a time dependent screening field and $\varepsilon_h$ is a time dependent thermal contribution to the strain. Using the experimental data, we have $\gamma$ ~220±75 for the 35 nm film and $\gamma$~160±50 for the 20 nm film (see SI Table S1). The errors are due to the uncertainty in determining the thermal contribution and to the fit itself. These values for $\gamma$ are higher than $\gamma$=40±60 inferred from the data between STO and (LaAlO$_3$)$_{0.3}$(Sr$_2$AlTaO$_3$)$_{0.7}$ (LSAT) substrates from Daumont et al. [24], who have measured a strong $d_{33}$ dependence of Mn-doped BFO films on the static epitaxial strain. However, these $\gamma$ value lie within two standard deviations of uncertainty of each other due to the comparably large error in each estimate. Given the differences in samples and measurement method, the difference is not unexpected.

The screening field $E_{\text{scr}}$ is time-dependent but homogeneous along the film depth, supplied by free charge carriers confined at the surface and the interface. The spatial homogeneity further requires negligible presence of free carriers within the film, which in turn indicates that laser-excited carriers are dominantly charge-neutral entities, i.e., bonded electron-hole pairs or excitons. These excitons dissociate and generate free carriers because of local band bending[25,26] when they diffuse to the surface and interface. The free carriers then either stay or migrate to the other side of the film depending on their charge sign and local polarization. During migration, due to their small concentration, the impact of these free carriers on the screening field is small but may be important at earlier times when the free carrier concentration is relatively high. However, direct separation of the charge carriers within the film by the internal polarization field can cause a dramatic field distortion that does not follow the static strain profile (see SI, Figs. S5 and S6).

This model, based on exciton dynamics mediated screening, is consistent with all experimental observations. The photoexcitation naturally steepens the existing strain gradient via dependence of the piezoelectric coefficient on the strain. As the screening field is inversely proportional to the film thickness, the steepening effect is more pronounced for thinner films thus increasing the broadening $\Delta w$. Dissociation of excitons located at or adjacent to the surface and interface lead to the instantaneous onset of the structural change[17] while the carriers inside the film must diffuse to the surface and interface to dissociate, leading to thickness-dependent dynamics. A schematic of this dynamic process is shown in Fig. 5.

Formation of such excitons in many perovskite oxides, including $PbTiO_3$, $BaTiO_3$, $KNbO_3$, and $KTaO_3$, has been discussed[27] in terms of the coupling of electron-hole pairs to the shift of the B-site ions and the neighboring oxygen atoms in the unit cell. This pair-coupled distortion forms in-gap energy states that smear the absorption edge[28] but do not lead to observable lattice effects. The exciton formation is characterized by distinctive absorption or photoluminescence spectral structures in the green, comparable to the absorption band between 500 and 600 nm (Fig. 3 (a)) in our experiment.

The band bending and exciton dissociation create depletion regions near the surface and interface. The field distribution in the depletion region can be quite different from that in the film and may be the reason that the structural effect does not exactly follow the expected scaling to the inverse of the thickness for the thinner films. Our current experiment does not provide the needed spatial resolution to reveal these details.

The giant enhancement of the strain gradient in ferroelectrics has significant practical implications. For BFO thin films, in which the static strain gradient is on the order of $2\times10^5$ m$^{-1}$, the steady state flexoelectric field is estimated to be 9 MV m$^{-1}$, which in large part determines the polarization of as grown films[8]. Doubling or tripling the strain gradient will increase the flexoelectric field to approximately 20 MV m$^{-1}$, close to the coercive field for most epitaxial ferroelectric thin films[30]. Applications such as direct optical writing of non-volatile ferroelectric memory[18] are thus now possible using dynamically induced flexoelectric polarization. Furthermore, as the enhancement is realized via a piezoelectric effect, it is therefore possible to manipulate the flexoelectric effect without the need of physical deformation, opening new horizons for dynamic flexoelectric effects in nanoscale devices.

This experiment is effectively also an electrode-free photovoltaic experiment that maps the field distribution and the charge carrier dynamics. The derived exciton-based model thus provides a new insight for ferroelectric photovoltaic effects. We also expect the time-dependent coherence analysis method to find wider applications in understanding complex correlations between structural and other degrees of freedoms in epitaxial thin films of other strongly correlated oxides.

## Methods

The transient optical absorption (TAS) experiment measures optical absorption of the sample as a function of the delay between a 40-fs, 400-nm pump laser pulse and the chirped 1 ps white-light probe laser pulse. The measured absorption spectrum has a wavelength ranging from 400 to 750 nm with time delays up to 7.2 ns. Time-resolved X-ray diffraction (TRXRD) experiments were performed at beam line 7ID-C of the Advanced Photon Source. Optical excitation was provided by 50 fs laser pulses with a central wavelength of 400 nm, derived by frequency-doubling the output of a Ti:sapphire laser system that was synchronized to the X-ray pulses with an electronically adjustable time delay. Incident X-ray pulses with photon energies of 10 or 12 keV and pulse duration of 100 ps were used for the series of TRXRD experiments described. Both the optical and the X-ray measurements were performed under ambient conditions with the pump spot size bigger than the probe by a significant margin. Other than a linear dependence of the amplitude, both the TAS and XRD dynamics were independent of the pump fluence up to the maximum used in the experiment at about 5 mJ/cm$^2$.

We used phase-pure epitaxial (001)-oriented BFO thin films of 4, 20, and 35 nm thickness grown on SrTiO$_3$ (STO) and (LaAlO$_3$)$_{0.3}$(Sr$_2$AlTaO$_3$)$_{0.7}$ (LSAT) substrates by reactive molecular-beam epitaxy[16]. The high quality of the samples were verified by X-ray reciprocal space mapping[20] from which the domain sizes are estimated to be about 400, 40, and 25 nm for the 4, 20, and 35 nm thick films, respectively. The pump photon energy (400 nm, 3.1 eV) was below the band gaps of STO (3.2 eV) and LSAT (5.2 eV), thus the photoresponse of the substrates were negligible in the measurement.

**Acknowledgement:** Work at Argonne was supported by the U.S Department of Energy, Office of Science, Office of Basic Energy Sciences, under Contract No. DE-AC02-06CH11357, and by American Recovery and Reinvestment Act (ARRA) funding through the Office of

Advanced Scientific Computing Research under Contract No. DE-AC02-06CH11357. Work at Cornell University was supported by Army Research Office through Agreement No. W911NF-08-2-0032. Work at the University of Wisconsin was supported by the U.S. Department of Energy, Office of Basic Energy Sciences, Division of Materials Sciences and Engineering, through Grant No. DEFG02-10ER46147.

**Author contributions:** YL conceived and designed the experiment, YL, PC, PGE, HW, DW and QZ performed or participated in the X-ray experiments and discussion, CR and RDS have performed the optical TAS experiment, CA grew the samples with advice from DGS. SN and WP performed the DFT simulations. YL wrote the manuscript, and all authors read and provided feedback on it. Correspondence should be sent to Yuelin Li: ylli@aps.anl.gov

Figures and Captions

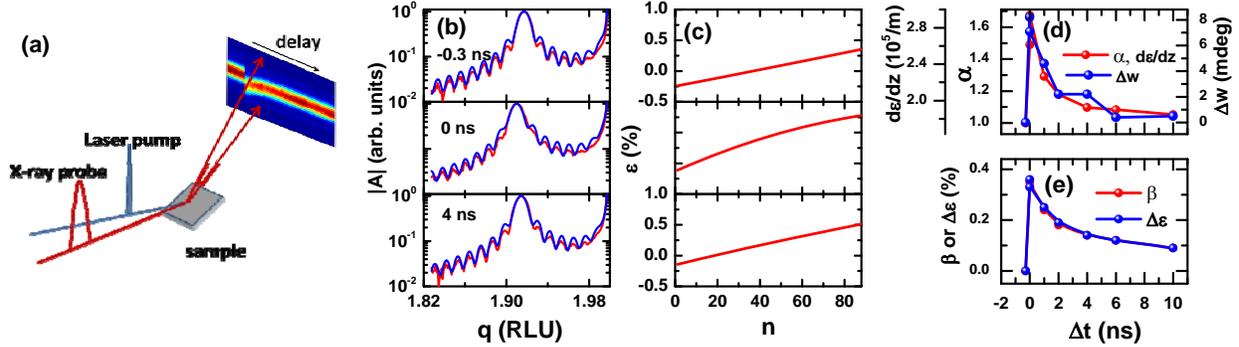

Figure 1. (a) Schematic of the time-resolved X-ray diffraction experiment using time-dependent coherence analysis. (b) Diffraction amplitude |A| along the coherent Bragg diffraction rod cut through the (002) diffraction peak for a 35 nm BFO thin film at a nominal fluence of 3.3 mJ/cm$^2$ for delays of -0.3, 0, and 4 ns (red) experimental data, (blue) fit. (c) Corresponding strain as a function of the monolayer index reconstructed via coherence analysis. More data are in Figs. S1-S3. (d) Parameters $\alpha$ and the corresponding strain gradient $d\varepsilon/dz$ in Eqs.(1) as a function of time $\Delta t$ (red) in comparison with measured width change $\Delta w$ (blue). (e) Parameter $\beta$ in Eq. (1) (red) and the measured average strain change $\Delta\varepsilon$ (blue). Note that, the broadening and the shift of the diffraction peak is a manifestation of the de-coherence and average phase shift of the diffracting unit cells. RLU: reciprocal lattice unit relative to that of STO.

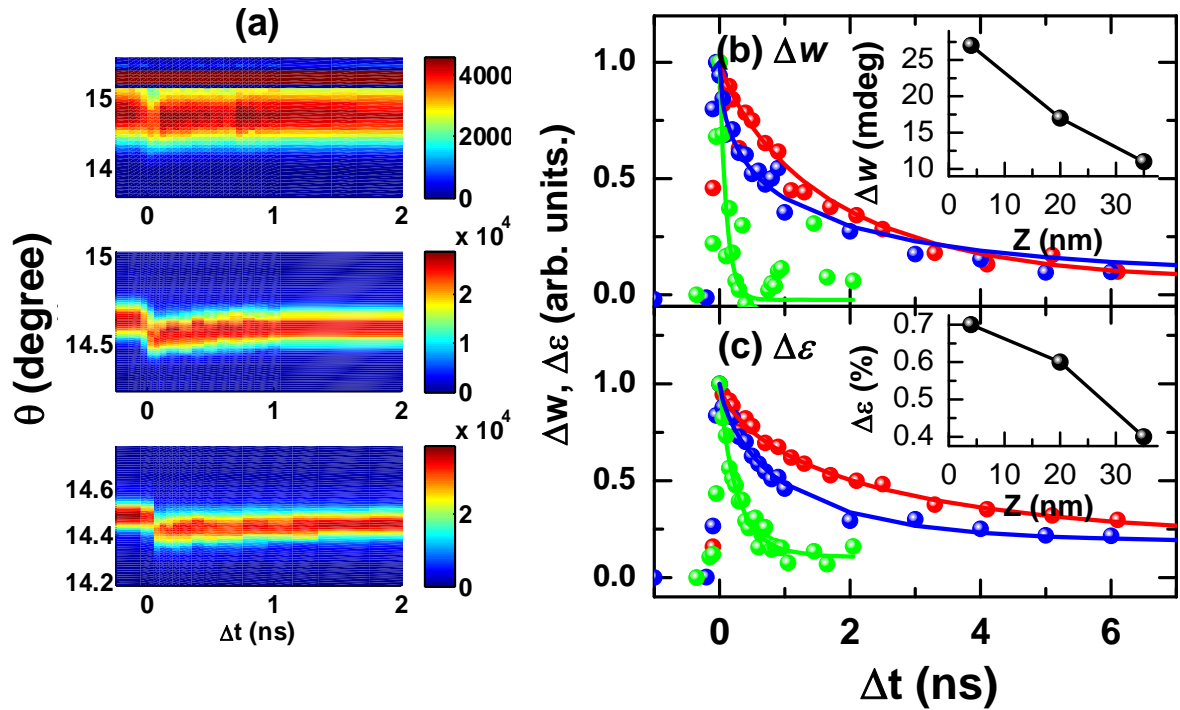

Figure 2 (a) X-ray diffraction near the 002 Bragg peak in BFO films as a function of $\Delta t$ with film thicknesses $Z$ = 4 nm (top), 20 nm (center) and 35 nm (bottom), taken at a nominal laser fluence of 2.5 mJ cm$^{-2}$. (b) Normalized broadening of the diffraction peak $\Delta w$ and (c) average strain $\Delta\varepsilon$ for $Z$= 4 (green), 20 (blue), and 35 nm (red), with their corresponding peak value at $\Delta t$=0 as a function $Z$ in the inserts. Solid lines are stretched exponential fits to the strain (see Fig. 4 caption). The intense feature at 15.2° in the top panel in (a) is the 002 Bragg peak from the STO substrate.

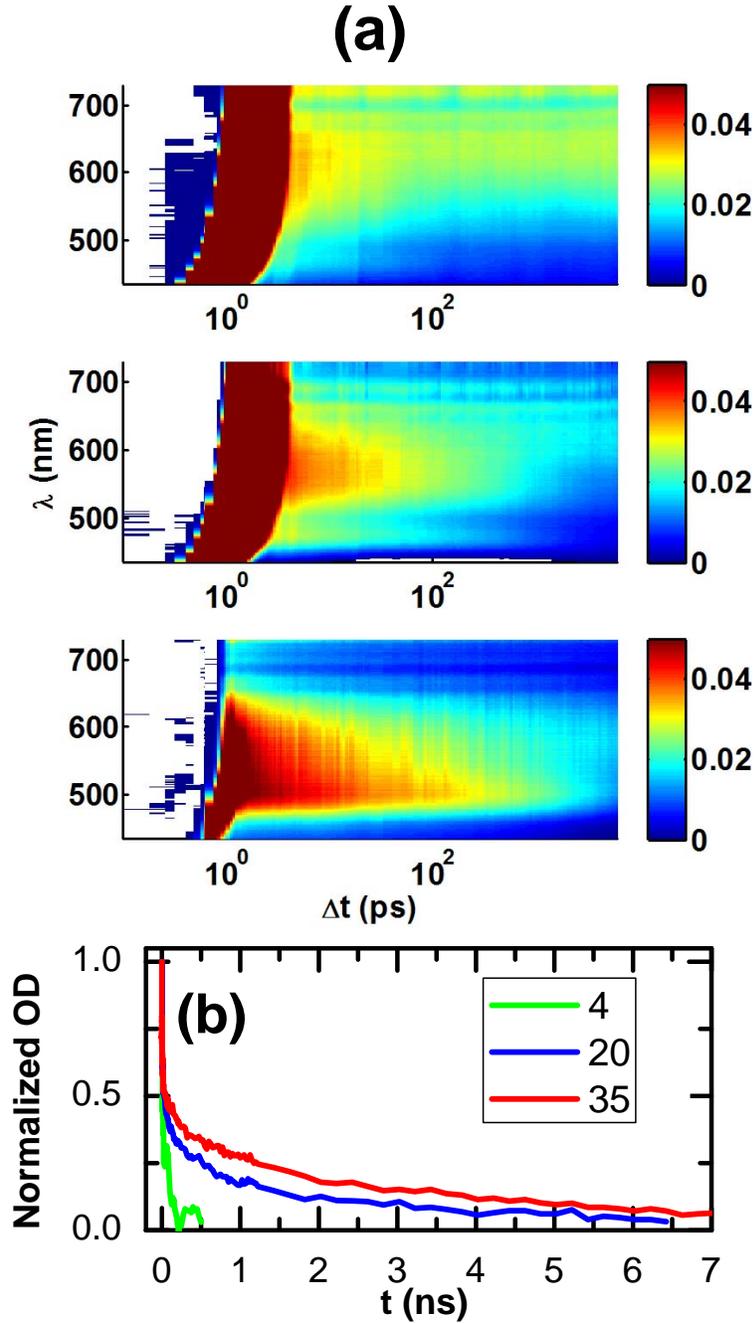

Figure 3 (a) Net change of the absorption spectra as a function of delay (Δ*t*) between the pump and the probe taken at nominal fluence of 5.5, 5.5, and 4.7 mJ/cm$^2$ for (from top to bottom) Z=4, 20 and 35 nm films. (b) OD as a function of delay for the three film thickness.

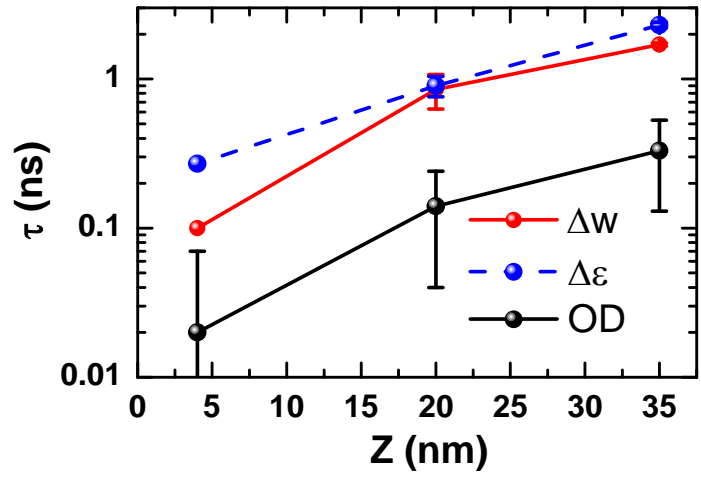

Figure 4 Recovery time for the relaxation of optical density (OD) and average strain ($\Delta\varepsilon$) and diffraction peak broadening ($\Delta w$). The 1/e recovery time $\tau$ is extracted by fitting to a stretched exponential function, $f(t)=a+b\exp(-(t/\tau)^\eta)$, with $a$, $b$ and $\eta$ being fitting parameters.

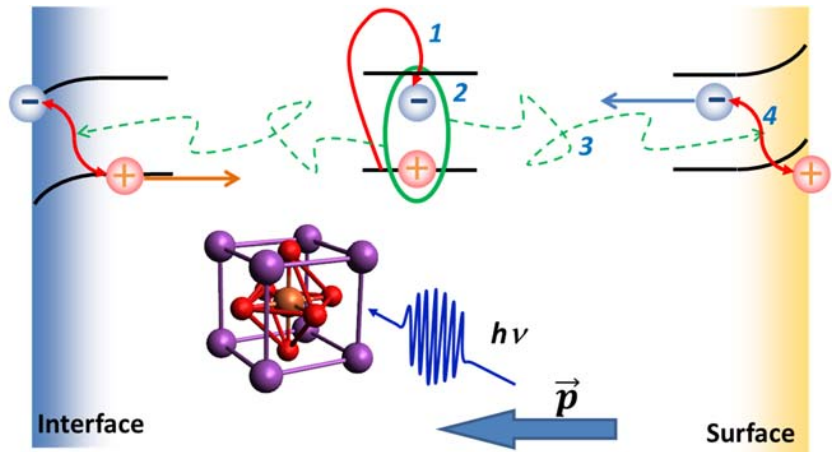

Figure 5 Schematic of the carrier dynamics leading to the steepening of the strain gradient. (1) electrons are excited and (2) cooled to form excitons. (3) Excitons then diffuse to the interface and surface where they (4) dissociate due to local band bending. The free carriers screen the depolarization field leading to a structural piezoelectric response. The strain dependence of the piezoelectric coefficient leads to the steepening of the static strain gradient. A BFO unit cell is shown in the inset with Bi, Fe, and O atoms represented by purple, brown, and red balls. Note that the initial film polarization direction has minimum effect on the lattice change.

# Giant optical enhancement of strain gradient in ferroelectric thin films and its physics origin

Yuelin Li[1], Carolina Adamo[2], Pice Chen[3], Paul G. Evans[3],

Serge M. Nakhmanson[4], William Parker[5], Clare E. Rowland[6], Richard D. Schaller[7],

Darrell G. Schlom[8,9], Donald A. Walko[1], Haidan Wen[1], and Qingteng Zhang[3]

[1] *Advanced Photon Source, Argonne National Laboratory, Argonne, Illinois 60439, USA*

[2] *Department of Applied Physics, Stanford University, Stanford, CA 94305, USA*

[3] *Department of Materials Science and Engineering & Materials Science Program, University of Wisconsin–Madison, Madison, Wisconsin 53706, USA*

[4] *Department of Materials Science & Engineering, and Institute of Materials Science University of Connecticut, Storrs, CT 06269-3136, USA*

[5] *Argonne Leadership Computing Facility, Argonne National Laboratory, Argonne, Illinois 60439, USA*

[6] *Department of Chemistry, Northwestern University, Evanston, IL 60208, USA*

[7] *Center for Nanoscale Materials, Argonne National Laboratory, Argonne, Illinois 60439, USA*

[8] *Department of Materials Science and Engineering, Cornell University, Ithaca, New York 14853, USA*

[9] *Kavli Institute at Cornell for Nanoscale Science, Ithaca, New York 14853, USA*

---

[1] Correspondence should be sent to Yuelin Li: ylli@aps.anl.gov

**Coherence diffraction analysis and strain profile retrieval methods**

The diffraction amplitude from a thin film with the contribution from the substrate is calculated using a kinetic diffraction model,

$$\left|A_{fit}(q)\right| = \left|FFT[\exp(i\varphi_n)]F_{BFO}(q) + \frac{aF_{STO}(q)}{[1-\exp(iq)]}\right|, \quad (s1)$$

Here $\varphi$ is the layer-by-layer BFO phase shift of the unit cells, $F_{BFO}$ and $F_{STO}$ are the structure factors, $n=1,..., N$ and is the layer index and $N$ is the number of epitaxial layers, and $a$ is the fitting parameter for the relative intensity of the substrate and the film, respectively.

The fitting of the strain profile takes our knowledge of the film thickness and is accomplished by a spline algorithm[1] using 4 evenly distributed fitting points for adjusting the phase while using a spline interpolation to obtain the phase at each unit cell layer. Fitting with more points does not change the fitting qualitatively and only reduce the fitting error modestly by less than 10%. The fitting algorithm tries to minimize the following error while changing the phase:

$$\sum_q ||A_{fit}(q)| - |A_{mea}(q)||, \quad (s2)$$

where $|A_{mea}(q)|=|I(q)|^{1/2}$ is the measured diffraction amplitude along the truncation rod.

The measurement was repeated at different delays for samples with different laser fluences and film thicknesses. A complete set of data is depicted in Figs. S1-S3. The fitting matches the position and the relative amplitude of the fringes well except for the 20 nm film at time zero, which needs further study to determine the cause.

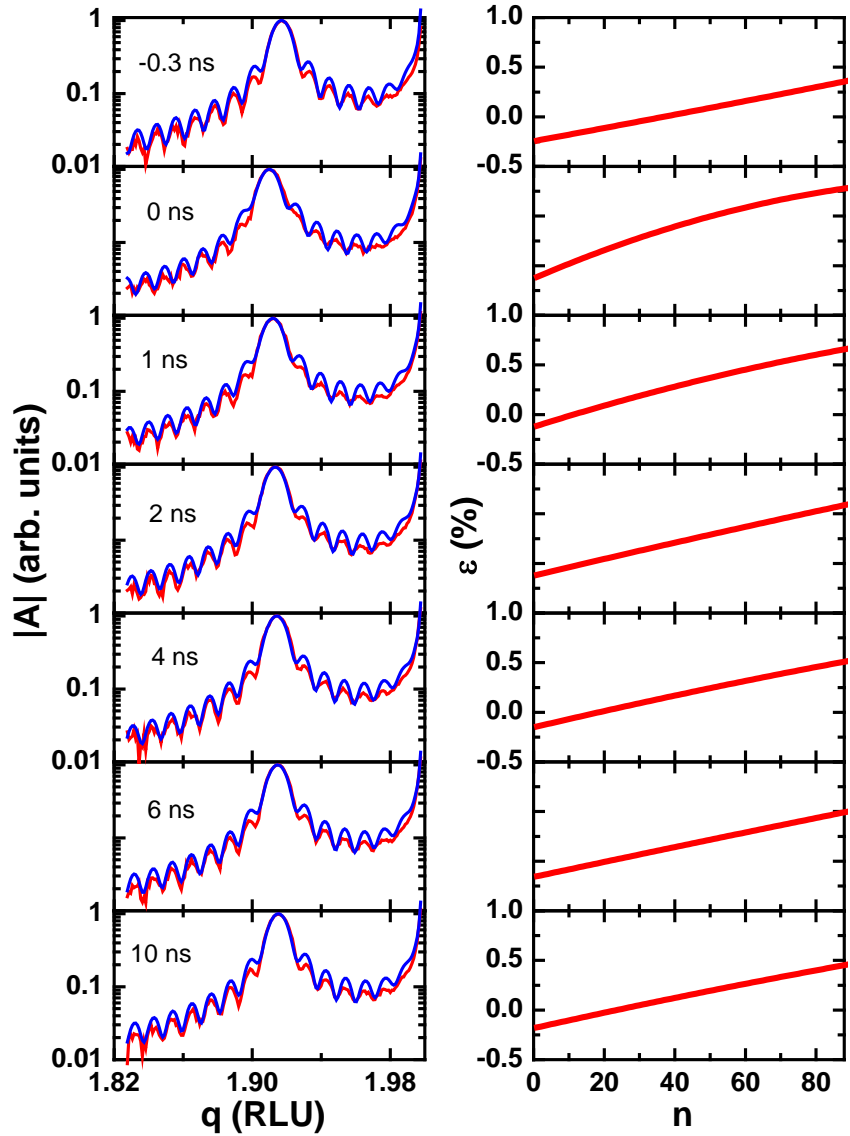

**Figure S1** A complete data set of the mapping of the strain as a function of delay for a 35 nm film (88 monolayers) at 3.3 mJ/cm$^2$. Left column: measured (red lines) and fitted (blue lines) diffraction amplitude |A| at different delays between the laser and the X-ray. Left column: corresponding fitted strain at different delays as a function of the film layer index *n*.

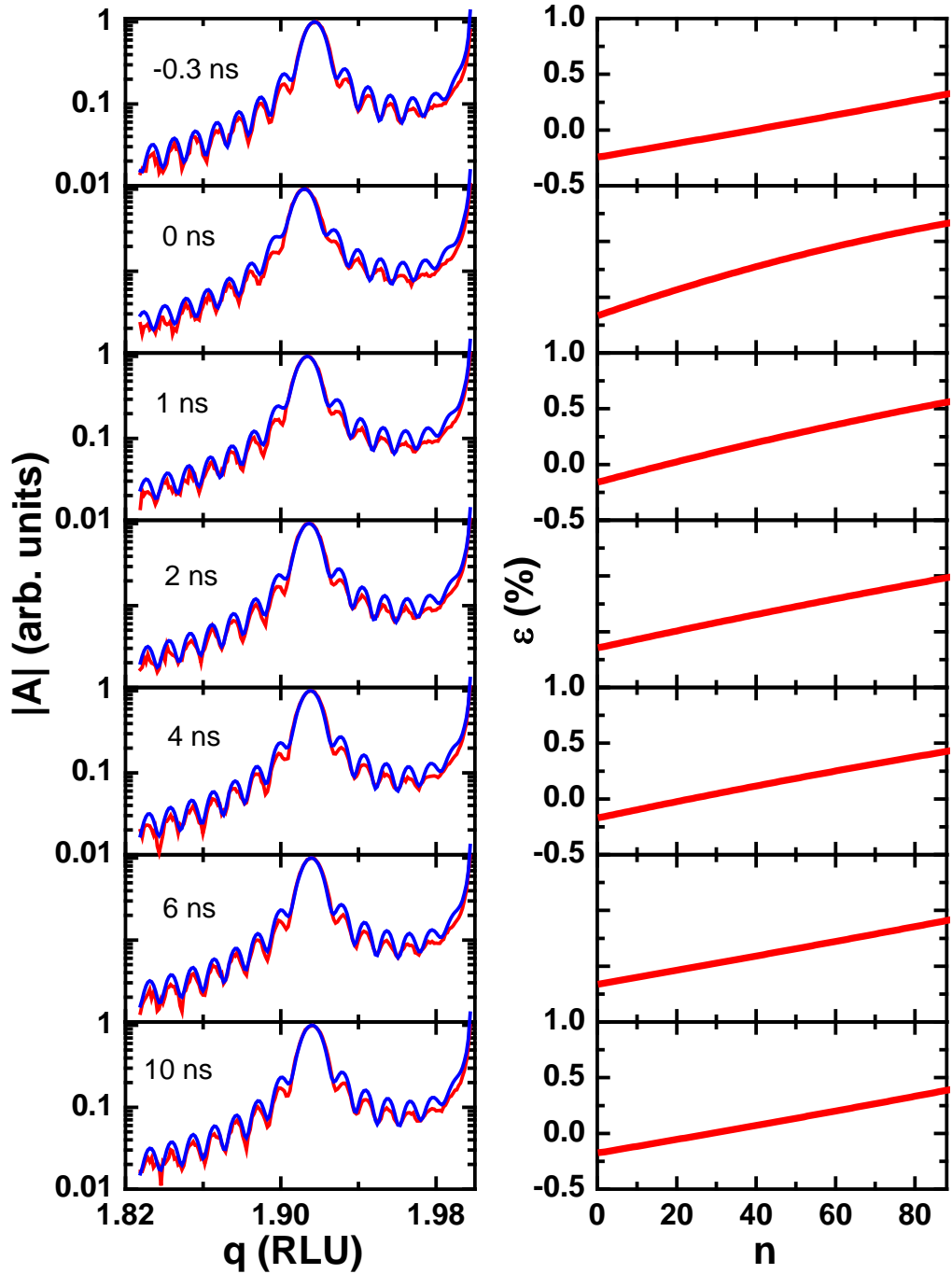

**Figure S2** (a) A complete data set of the strain as a function of delay for a 35 nm film (88 monolayers) at 2.5 mJ/cm$^2$. Left column: measured (red lines) and fitted (blue lines) diffraction amplitude |A| at different delays between the laser and the X-ray. Left column: corresponding fitted strain at different delays as a function of the film layer index n.

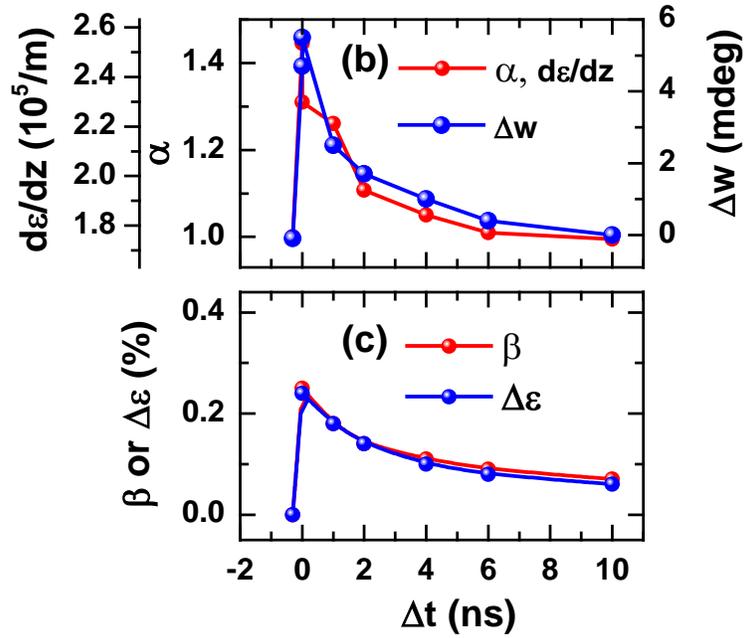

**Figure S2** (b) Fitting parameters $\alpha(t)$ and the corresponding strain gradient $d\varepsilon/dz$ in Eqs.(1) (red) in comparison with measured width change $\Delta w(t)$ (blue). (c) Parameter $\beta(t)$ in Eq. (1) (red) and average strain change $\Delta\varepsilon(t)$ (blue).

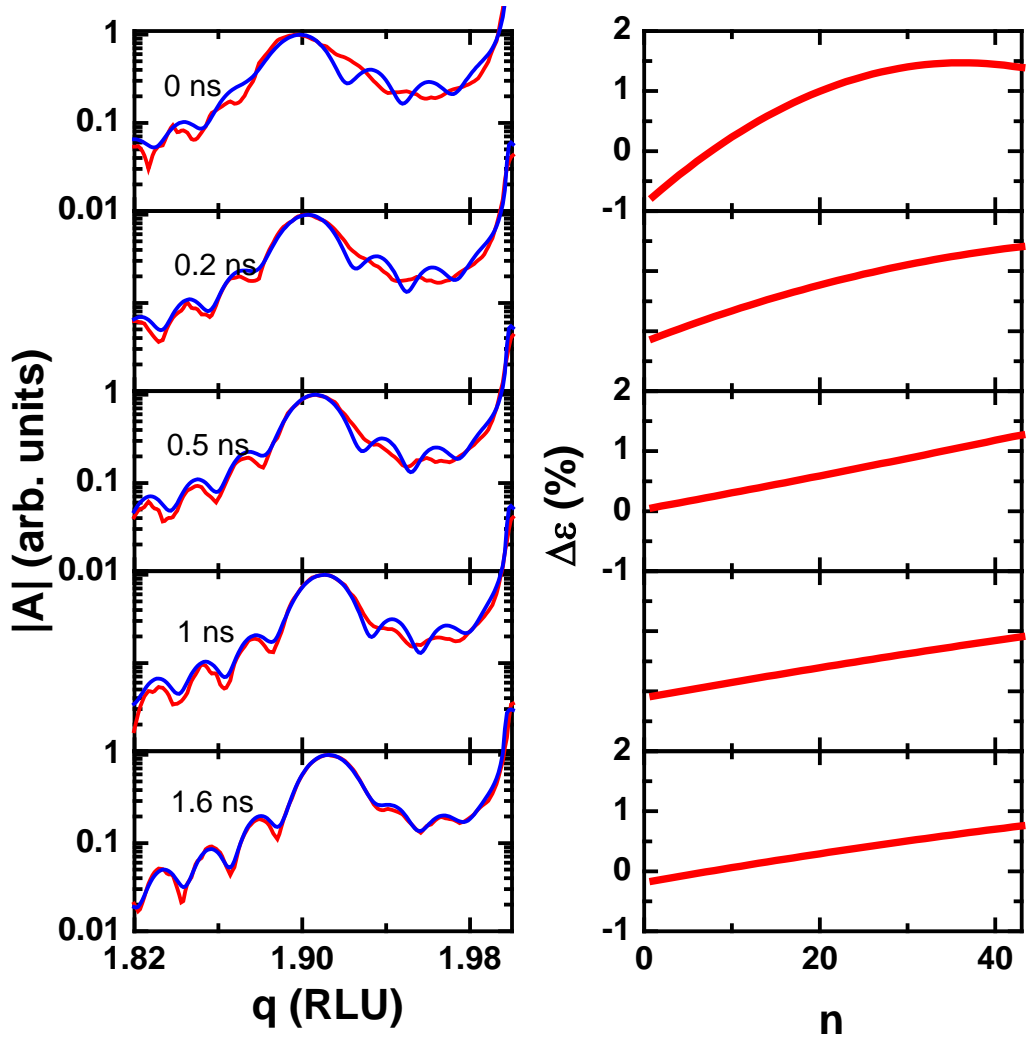

**Figure S3** (a) A complete data set of the mapping of the strain as a function of delay for a nominal 20 nm film (50 monolayers) at 3.3 mJ/cm$^2$. Left column: measured (red lines) and fitted (blue lines) diffraction amplitude |A| at different delays between the laser and the X-ray. Left column: corresponding fitted strain at different delays as a function of the film layer index n. The fitting is very good for all delays except for zero delay.

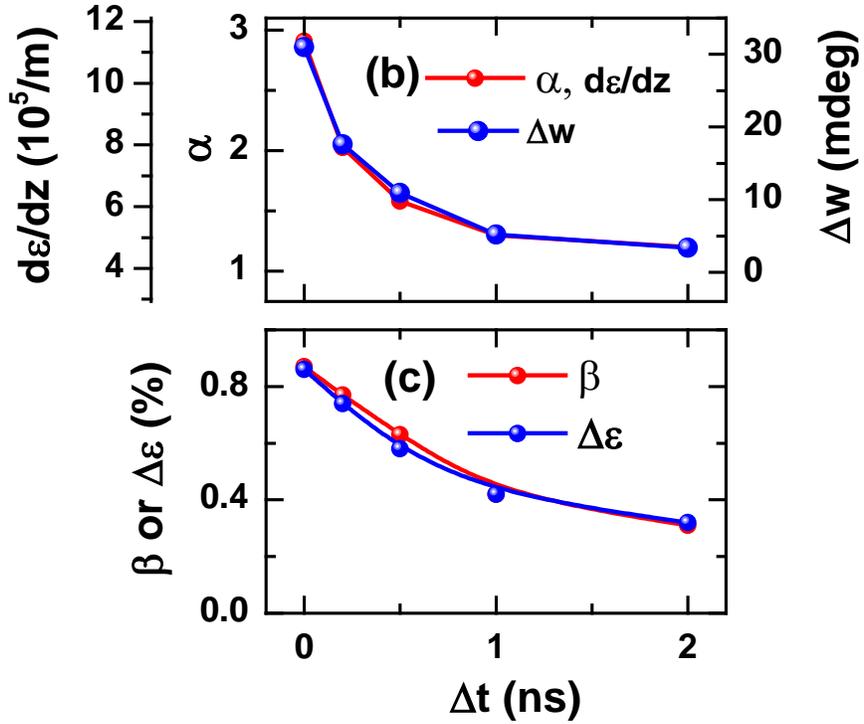

**Figure S3** (b) Fitting parameters $\alpha(t)$ and the corresponding strain gradient $d\varepsilon/dz$ in Eqs.(1) (red) in comparison with measured width change $\Delta w(t)$ (blue). (c) Parameter $\beta(t)$ in Eq. (1) (red) and average strain change $\Delta\varepsilon(t)$ (blue).

**Supplementary Information 2**

**DFT simulation**

DFT calculations were performed in QUANTUM ESPRESSO[2] using the local density spin approximation[3,4] and an on-site Coulomb parameter[5] of $U = 4$ eV applied to the Fe states[6]. The Fe atoms were initialized to an antiferromagnetic spin configuration. Vanderbilt ultrasoft pseudopotentials[7] with scalar relativistic corrections simulated the core and valence electrons. The pseudopotentials were generated with the Perdew-Zunger parameterization[8] of the local density approximation in DFT using the following parameters: Bi: $6s^2 5d^{10} 6p^3$, $r_0$=1.2 bohr, $r_{loc}$=2.2 bohr, $r_c$ = (2.5,2.5,2.2) bohr for $s$, $p$, and $d$, respectively. Fe: $3s^2 3p^6 4s^2 3d^6 4p^0$, $r_0$=1.5 bohr, $r_{loc}$=2.0 bohr, $r_c$ = (2.0,2.0,2.0) bohr for $s$, $p$, and $d$, respectively. O: $2s^2 2p^4$, $r_0$=0.7 bohr, $r_{loc}$=1.0 bohr, $r_c$ = (1.2,1.2) bohr for $s$ and $p$, respectively. A plane wave basis supporting the wave function (density) cutoff at 50 Ry (400 Ry) converged the rhombohedral $R3c$ and cubic perovskite $Pm\bar{3}m$ structural energy difference to within 1 meV per formula unit. A Γ-centered 5×5×5 Monkhorst-Pack k-point mesh for the $R3c$ structure sampled the Brillouin zone to converge total energy to within 10 meV per function unit. The sum of the forces on the ions was relaxed in the $R3c$ symmetry to less than 20 meV/angstrom, and the pressure on the simulation cell in the same symmetry to less than 0.01 kbar. A Fermi-Dirac distribution applied to the occupation of the DFT single-particle states[9] simulated the effect of excitation by varying the width of the distribution from 0.00 eV up to 2.04 eV to increase electron temperature. The gap between the highest occupied state and the lowest unoccupied DFT single-particle state disappears between Fermi-Dirac distribution widths 0.95 and 1.08 eV. The relaxed spin state transitions from antiferromagnetic to nonmagnetic between Fermi-Dirac distribution widths of 1.36 and 1.50 eV. Bi and Fe ions retain a presumably ferroelectric displacement from centrosymmetric positions with $R\bar{3}c$ symmetry at all Fermi-Dirac distribution widths simulated.

The rhombohedral lattice parameters were then converted to pseudo-cubic. The out-of-plane strain was calculated according to Hooke's law, taking into account the coupling of epitaxial in-plane stress due to the cold substrate:

$$\varepsilon = \frac{2\nu}{1-\nu}\varepsilon_{in}, \quad (s3)$$

where $\nu = 0.34$ is the Poisson ratio[10], and $\varepsilon$ and $\varepsilon_{in}$ are the out-of-plane and in-plane strain, respectively.

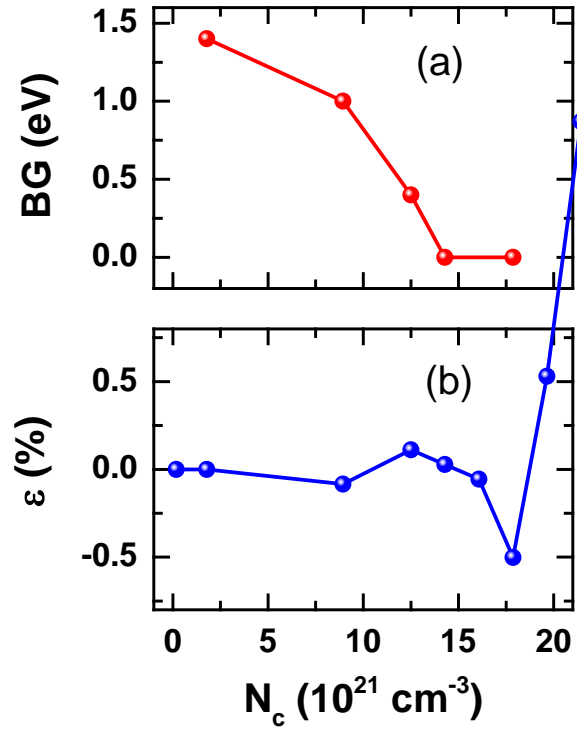

**Figure S4** Kohn-Sham band gap (BG) (a) and out-of-plane strain (b) as a function of the carrier density ($N_c$) from DFT simulations. The strain calculation takes into account the Poisson effect due to the in-plane stress. The $N_c$ corresponding to the highest pump fluence of 5 mJ cm$^{-2}$ in the experiment is about $1.5\times10^{21}$ cm$^{-3}$.

**Supplementary Information 3**

**Dependence of the piezoelectric coefficient on the strain**

From Eqs. (2 -4), we have

$$\alpha(t) - 1 = \gamma d_0 E_{scr}(t). \tag{s4a}$$

$$\gamma = \frac{\alpha(t)-1}{E_{scr}d_0} = \frac{\alpha(t)-1}{\beta(t)-\varepsilon_{scr}(t)}. \tag{s4b}$$

The resulting $\gamma$ using our experimental data is shown in Table S1.

**Table S1. Parameter $\gamma$ obtained from the experimental data. The errors are due to the uncertainty in determining the thermal contribution and the uncertainty in the fitting.**

|  | 20 nm, 3.3 mJ cm$^{-1}$ | 35 nm, 3.3 mJ cm$^{-1}$ | 35 nm, 2.5 mJ cm$^{-1}$ |
|---|---|---|---|
| Delay (ns) | 0.2 | 0 | 0 |
| $\alpha$ | 1.97 | 1.57 | 1.38 |
| $\beta$ | 0.76% | 0.37% | 0.26% |
| $\varepsilon_h$ | 0.16% | $\beta$/3 | $\beta$/3 |
| $\gamma$ | 160±50 | 234±80 | 216±70 |

**Supplementary Information 4**

**Disproving the charge separation by polarization field model**

Bulk charge separation due to internal polarization or external field leads to strong distortion of the field in a pump probe experiment though it may have no effect in a CW experiment. We simulated the situation with a one dimensional particle-in-cell dynamic model for the 35 nm film where the motion of the carriers (Fig. S4(a)) and the field (Fig. S4(b)) are solved self-consistently. Equal number of holes and electrons are generated filling the space with a profiled probability $p(z)=\exp(-z/Z)$, where $Z = 32$ nm is the absorption length at 400 nm. To illustrate the physics, we apply a constant field $E_0 = 1$ MV/cm that generates a peak strain of 0.5% at a nominal piezoelectric constant $d_{33} = 50$ pm/V. We use nominal mobility for the electrons and holes of $7\times10^{-5}$ and $5\times10^{-5}$ m$^2$V$^{-1}$s$^{-1}$, respectively, for illustrative purpose. The result is not sensitive to the choice of these numbers other than the time scale. A dielectric constant of 50 is used[11]. The carriers are absorbed when they reach the boundaries. We also consider only the cases where carriers are immediately separated at birth.

For carrier density higher than $3\times10^{18}$ cm$^{-3}$ (corresponding roughly to an absorption fluence of 0.01 mJ/cm$^2$), a $\pm100\%$ modulation of the field can be achieved, indicating the saturation of the applied field. For lower carrier densities, the maximum modulation is proportional to the carrier density. Fig. S3 shows a case with carrier density of $1.5\times10^{18}$ cm$^{-3}$ (absorption fluence of 0.005 mJ/cm$^2$) As can be seen in Figure S3 (b), the charge separation induces a $\pm50\%$ modulation of the field inside the film. When mapped into the piezoelectric response of the unit cells, the modulation will lead to a strain profile completely different from that arises from a uniform field expected from the exciton scenario where carriers only separate at the surface and the interface. The low carrier density needed for such modulation also demonstrates the sensitivity of the strain profile to such a bulk space charge effect. Simulation using strain profile following the field profiles with comparable strain range as measured from the experiment generates significantly asymmetric fringes around the central diffraction peak, which is not observed in the experimental data (Fig. S5), disproving the commonly accepted carrier separation by polarization field model.

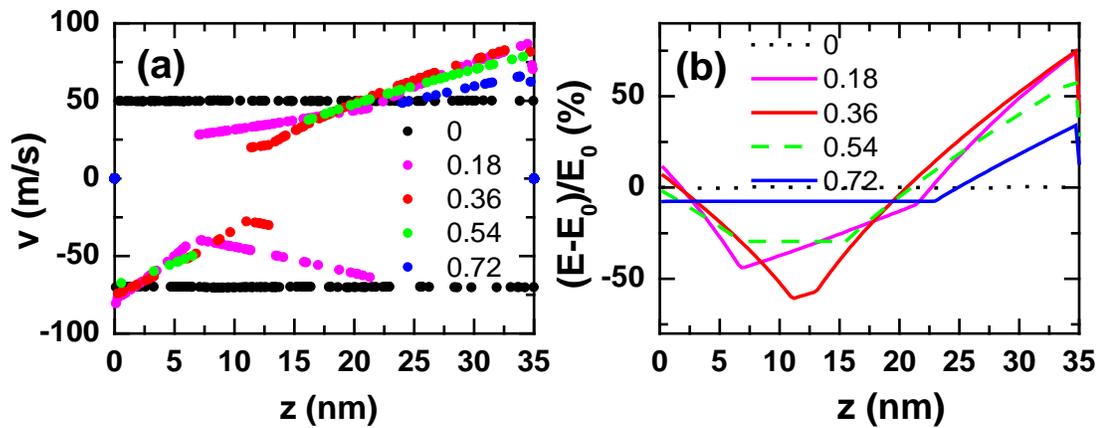

**Figure S5** Effect of the field inside a 35 nm film when the carriers are generated by an impulsive excitation and immediately separated by an external field. (a) Position-velocity phase diagram of the holes (positive speed) and electrons (negative speed) at different times (indicated by the legend in ns) and (b) the corresponding field modulation as the carriers are separated by the applied field of $E_0=1$ MV/cm. The initial carrier density is $1.5\times10^{18}$ cm$^{-3}$ corresponds roughly to an absorption fluence of about 0.005 mJ/cm$^2$, less than 1% of the that used in our experiment. The asymmetry in the field and phase diagram are due to the difference in the electron and hole mobility.

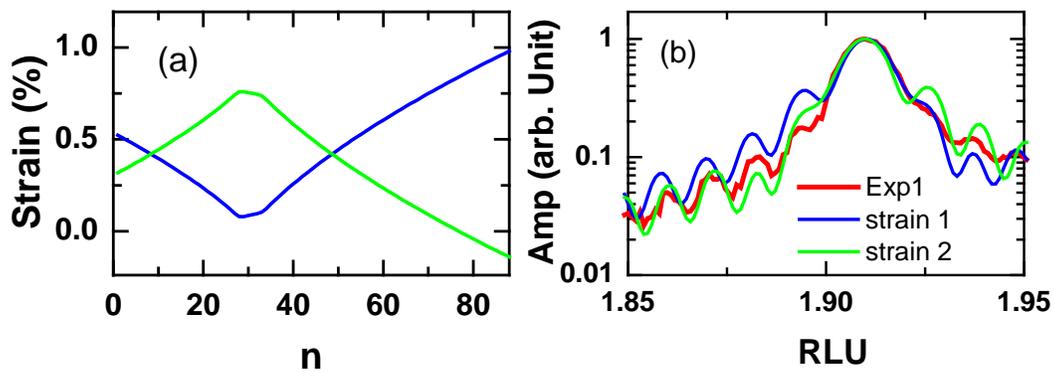

**Figure S6** Simulation of the diffraction pattern using strain profile derived from the field distribution in Fig. S4. (a) Two possible strain profiles and (b) the simulated diffraction patterns near the 0 0 2 diffraction peak in comparison with the experiment measurement for the 35 nm high fluence case. To obtain the strain profile, we map the field profile to the strain range observed in the experiment with a shift to match the diffraction peak position. A strong fringe intensity asymmetry is predicted but not observed in the experiment. This confirms the quality of our fit in Figs. 1, S1-S3 and the interpretation of the data, disproving the commonly accepted carrier separation by polarization field model.

`